\documentclass[reprint,superscriptaddress,nobibnotes,amsmath,amssymb,aps,prm,floatfix,showkeys]{revtex4-2}
\usepackage{xr}
\usepackage{filecontents}
\usepackage{graphicx}
\usepackage{amsmath}
\usepackage{color}
\usepackage{xcolor}
\usepackage{subfiles}
\usepackage{float}
\usepackage{hyperref}
\hypersetup{colorlinks=true, linkcolor=blue, filecolor=blue, urlcolor=blue, citecolor=blue }
\begin{document}

\title{The origin of phase separation in binary aluminosilicate glasses}

\author{Houssam Kharouji}
\affiliation{LEM3,Labex Damas, Université de Lorraine, Metz, 57070, France}
\affiliation{LS2ME, Facult\'{e} Polydisciplinaire Khouribga, Sultan Moulay Slimane University of Beni Mellal, B.P 145, 25000 Khouribga, Morocco}
\author{Abdellatif Hasnaoui}
\affiliation{LS2ME, Facult\'{e} Polydisciplinaire Khouribga, Sultan Moulay Slimane University of Beni Mellal, B.P 145, 25000 Khouribga, Morocco}
\author{Achraf Atila}
\email{achraf.atila@uni-saarland.de; achraf.atila@gmail.com}
\affiliation{Department of Material Science and Engineering, Saarland University, Saarbr\"{u}cken, 66123, Germany}

\begin{abstract}
The quest for hard and tough transparent oxide glasses is at the core of glass science and technology. Aluminosilicate glasses exhibiting nanoscale phase separation emerge as promising candidates for such materials. Nevertheless, proper control of the phase separation represents a daunting challenge due to its elusive origins. Here we employ large-scale molecular dynamics simulations and structural analysis to unravel the underlying mechanisms of the phase separation in aluminosilicate. The observed phase separation originates from an arrangement of SiO$_4$ and AlO$_n$ polyhedra, which manifests from the second coordination shell and extends to higher shells. This specific arrangement is driven by repulsion between the polyhedra, reaching its maximum at around 50 mol\% of Al$_2$O$_3$. This behavior becomes pronounced around and below the glass transition temperature. This work sheds light on the origin of phase separation and provides a route for further exploration across other compositions to develop glasses with adapted mechanical performance.

\end{abstract}


\date{\today}
\maketitle 
Oxide glasses are used in a wide range of applications, enabling modern civilization, e.g., optical fibers~\cite{Wondraczek2022}, bioactive materials~\cite{atila2022atomistic, ouldhnini2021atomistic}, solar cells~\cite{Axinte2011}, and flexible and foldable screens for mobile devices~\cite{luo2015}. 
The usage of oxide glasses in these applications benefits from their excellent optical~\cite{Herrmann2023, Jagannath2022}, chemical durability~\cite{Xiang2013}, and the high tailorability of the properties through compositional design~\cite{atila2019atomistic, atila2022atomistic}, (post-) processing~\cite{Ganisetti2023, Bakhouch2024}, or a combination thereof~\cite{Smedskjaer2015}. 
However, the brittleness of these glasses, which is due to the lack of a stable shearing mechanism that accommodates the deformation, has limited the development of existing and emerging glass applications~\cite{Rouxel2017}. 
Thus, the need for new glasses with adapted mechanical properties triggered efforts to achieve this goal~\cite{Atila2024, Pan2024, Zhang2021, Januchta2017jncs, Januchta2017chem, Rosales2016}. 

For instance, the use of reinforcements has been proven effective in controlling crack propagation~\cite{Shi2014}, it affected the glass transparency~\cite{Yin2019}. Glasses with nanoscale phase separation were observed to improve fracture toughness compared to homogeneous glasses~\cite{Yuan2014, Tang2018}. 
The mechanism of enhancing the hardness and fracture toughness of glasses with phase separation works in principle in the same way as for a composite material with two interconnected glassy regions with different structures, i.e., a rigid one and the other one is more flexible~\cite{Januchta2017chem, Tang2018, Wang2016nano}. 

Binary aluminosilicate glasses with nanoscale phase separation are great potential candidates to make glasses with high hardness and crack resistance~\cite{Rosales2016, Xia2023}. 
Rosales-Sosa~\textit{et al.} studied aluminosilicate glasses with alumina content ranging between 30 and 60 mol\% and showed that the increase in the elastic moduli, hardness, and cracking resistance of binary aluminosilicate glasses was due to the increase in the atomic packing density, the high dissociation energy per unit volume of Al$_2$O$_3$ compared to that of SiO$_2$, and increase in the glass density with increasing Al$_2$O$_3$ content~\cite{Rosales2016}. 
While the explanations from Ref.~\cite{Rosales2016} are indeed valid, they neglect the presence of a phase separation, which shows a maximum at the compositions with an alumina content around 50 mol\%~\cite{macdowell1969, risbud1977, KLUG1987, Poe1992, Djuric1996, Wilke2022, Ohkubo2024}, and its effects on the observed behavior. 
Macdowell \textit{et al.} highlighted that the composition at which the alumina content is around 50 mol\% shows the highest degree of phase separation~\cite{macdowell1969}. While others showed that the maximum phase separation occurs around 60 mol\% of Al$_2$O$_3$ ~\cite{Wilke2022, Wilke2023}.

Although phase separation has been observed in binary~\cite{Hudon2002, Sen2004, Liao2020}, ternary~\cite{Martel2011} aluminosilicate, and other oxide glasses~\cite{Sreenivasan2020}, its origins remained elusive. 
Here we used a million atoms molecular dynamics (MD) simulations to study the phase separation in aluminosilicate glasses with Al$_2$O$_3$  content ranging between 10 and 90 mol\%. We found that this phase separation is discernible through changes in the medium-range structure, starting from the second coordination shell and extending to larger shells. Furthermore, our MD results indicate an absence of any significant phase separation in the liquid state at elevated temperatures, and signs of phase separation were observed only in the supercooled liquid around the glass transition temperature (T$_g$) and below it.

All MD simulations were performed using LAMMPS~\cite{Thompson2022}. The potential from Bouhadja \textit{et al.}~\cite{Bouhadja2013} was used to model the interactions between atoms. The glass composition of the binary aluminosilicate glasses has Al$_2$O$_3$ content ranging between 10~mol\% to 90~mol\%. The glass was equilibrated at 5000~K and cooled using 1~K/ps to room temperature (300~K). Configurations at different temperatures during the cooling process were extracted for further analysis. More details about the glass preparation can be found in Ref.~\cite{atila2019alumina} and the supplementary materials (SM). The visualizations are made using OVITO~\cite{Stukowski2009}. The structure of the glass is in realistic agreement with the data available in the literature~\cite{Bouhadja2013, Bauchy2014, weber2008} and as discussed in the SM and shown in Fig.~S1, Fig.~S2, and Tab.~S1.

\begin{figure}[h!]
\centering
\includegraphics[width=\columnwidth]{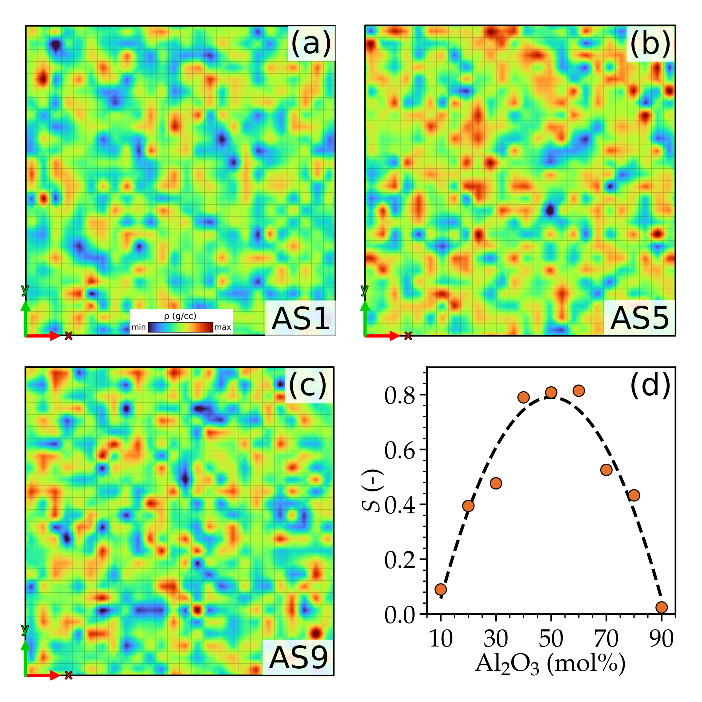}
\caption{(a - c) Selected local density maps of binary aluminosilicate glasses. The density maps are for (a) $x$ = 10 mol\%, (b) $x$ = 50 mol\%, and (c) $x$ = 90 mol\%. The simulation boxes were divided to produce small cubes of volume, 10$\times$10$\times$10 \text{\AA$^3$}, and then the density of each cube was calculated and used for further analysis. In (d), the skewness is plotted as a function of the composition at 300~K. The skewness is obtained by fitting the histograms of local density to a skewed Gaussian function (See Supplementary material for details). The dashed line is a fit to a quadratic function and serves as a guide to the eye (For interpretation of the colors in the figure(s), the reader is referred to the web version of this article). }
\label{Fig:densityskew}
\end{figure}

The phase separation was visually observed in the simulated aluminosilicate glasses with different degrees as illustrated from selected compositions in Fig.~\ref{Fig:densityskew}(a-c) using the 2D spatial distribution of the mass density. The histograms of the density distribution throughout the system are shown in the SM, Fig~S3, which were fitted using a skewed Gaussian distribution function (details in the SM) to extract the skewness $S$. If the glass is homogeneous, then the density distribution should be well described through a Gaussian function, with its skewness being nearly 0. On the other hand, if the glass has phase separation, the density distribution can no longer be described through a symmetric Gaussian function, and it will be skewed. The higher the values of the skewness, the higher the asymmetry of the distribution. The values of the skewness $S$ are plotted as a function of the alumina content and show a maximum at composition around 50 mol\% of Al$_2$O$_3$. 

The glasses are expected to be more homogeneous at the compositions at which the SiO$_2$ content is very high or very low. The presence of inverted parabolic-like behavior of $S$ with the alumina content and the occurrence of a maximum at around 50 mol\% of Al$_2$O$_3$ is consistent with the hypothesis stating that the maximum phase separation is around 50 mol\% of Al$_2$O$_3$. 
These results are in good agreement with experiments and theoretical studies, although there is disagreement regarding the value of the critical composition of 
demixing in binary AS glasses~\cite{risbud1977,Poe1992,Ban1996}. 

From an atomistic point of view, the tendency to phase separate in AS glasses could be correlated with the preference of SiO$_4$ and AlO$_n$ (with $n$ = 4,5, or 6) polyhedra to be surrounded by those of the same type. 
This process is evaluated with the distance-dependent Warren-Cowley (WC) order parameter defined for a pair of  atom types $i,j$ as follows~\cite{Cowley1950, Bakhouch2024}:
\begin{equation}
\label{eq:CRO}
\alpha_{ij}(r) = 1 - \frac{n_{ij}(r)}{N_i(r)c_j},
\end{equation}

where n$_{ij}(r)$ is the number of atoms of type $j$ neighbors of a central atom of type $i$ at a distance $r$, $N_i(r)$ is the total of neighbors for the atom of type $i$ at a distance $r$, c$_j$ is the concentration of type $j$.
Negative values of $\alpha_{ij}$ indicate attraction between the pairs, while positive values correspond to the opposite. Throughout this paper, we limited the calculation of $\alpha_{ij}$ to Si and Al atoms without involving oxygen atoms to extract the correlations between SiO$_4$ and AlO$_n$ polyhedra with $n$ = 4,5, or 6.
Fig.~S4 represents the variation of this order parameter in the first coordination shell, defined by the first minimum of the radial distribution function, as a function of Al$_2$O$_3$ content. The calculated $\alpha_{\rm SiSi}$ is slightly higher than zero up to $x$ = 0.6, where it starts to increase. At low Al$_2$O$_3$ content, $\alpha_{\rm SiAl}$ is negative and increases with increasing Al$_2$O$_3$ content. This indicates a preference for Si to have Al in their first neighboring shell, which decreases with increasing alumina. On the other hand, $\alpha_{\rm AlSi}$ is always positive, highlighting that AlO$_n$ polyhedra do not prefer SiO$_4$ tetrahedra as neighbors and indicating a relatively low probability of having SiO$_4$ tetrahedra around an AlO$_n$ polyhedra in the first shell. The WC parameter for Al-Al is negative at low alumina content and increases with increasing Al$_2$O$_3$. This indicates that there is a strong attraction between AlO$_n$ and AlO$_n$ polyhedra, and AlO$_n$ prefers to have other AlO$_n$ polyhedra in their first neighboring shell, which becomes less pronounced with increasing Al$_2$O$_3$. 

\begin{figure}[ht!]
\centering
\includegraphics[width=\columnwidth]{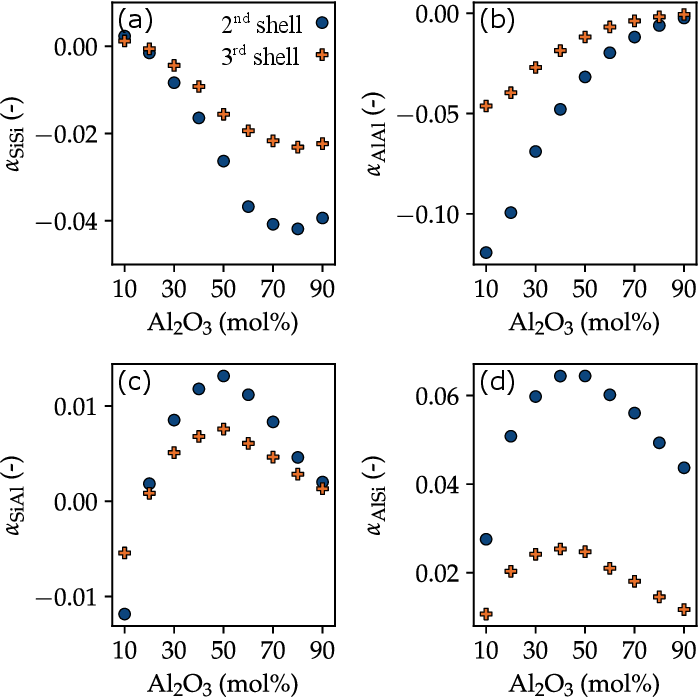}
\caption{Chemical short-range order at the second and third shell
as a function of the alumina content at 300~K. (a) Si-Si, (b) Al-Al, (c) Al-Si, and (d) Si-Al. (For interpretation of the colors in the figure(s), the reader is referred to the web version of this article).}
\label{Fig:SROhigherShells}
\end{figure}
Since the data obtained from the $\alpha_{ij}$ up to the first coordination shell, involving neighboring polyhedra, do not show clear signs of phase separation, which is expected as the phase separation spans over larger length scales. The investigation of the chemical arrangement of the glass building blocks at larger coordination shells is necessary. In Fig.~\ref{Fig:SROhigherShells}, the evolution of the WC parameters in the second and third coordination shells\footnote{We note that an exact determination of the higher coordination shells in amorphous materials is far from trivial. Thus, $r^{\rm 2^{nd}shell} = 5.35\rm~\AA$ and $r^{\rm 3^{rd}shell}~=~9.1\rm~\AA$. The values of the cutoff radii are provided for better reproducibility of the results.} of all considered compositions is represented. Comparing the results obtained from Fig.~S4, which focuses on the first shell exclusively, with those presented in Fig.~\ref{Fig:SROhigherShells}, a significant change in the WC parameter is observed for the Si-Si pair. $\alpha_{SiSi}$ calculated from, the second shell is positive at the composition with 90 mol\% of SiO$_2$ and becomes negative with increasing Al$_2$O$_3$ content. This indicates that the second and third shells of Si prefer to have Si than Al, which is the first indication of the phase separation.
The $\alpha_{AlAl}$ in the second and third shells remains negative and follows the same behavior as discussed earlier for the first shell. This finding indicates the preferred ordering between the AlO$_n$ polyhedra.  
More interestingly, the WC parameter for the pairs Si-Al and Al-Si shows a maximum around x = 0.5, consistent with the data provided in Fig.~\ref{Fig:densityskew}(d). A gradual increase is observed for both pairs up to a maximum and decreases again. This shows that a maximum in the repulsion between SiO$_4$ tetrahedra and AlO$_n$ polyhedra is around the composition at which R = SiO$_2$/Al$_2$O$_2$ = 1.

\begin{figure}[h!]
\centering
\includegraphics[width=\columnwidth]{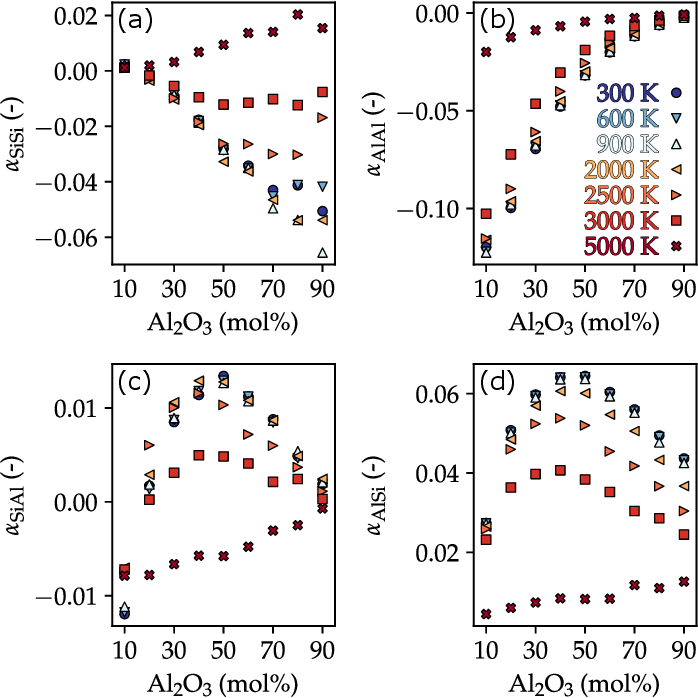}
\caption{Change of $\alpha_{ij}$ at the second shell as a function of the composition for different selected temperatures above and below $T$\textsubscript{g}. (a) is for $\alpha_{\rm SiSi}$, (b) is for $\alpha_{\rm All}$, (c) is for $\alpha_{\rm SiAl}$, and (d) is for $\alpha_{\rm AlSi}$. The data for all samples as a function of the temperature is shown in Fig.~S5. (For interpretation of the colors in the figure(s), the reader is referred to the web version of this article).}
\label{temp2ndshell}
\end{figure}

The temperature-dependent evolution of the WC parameter $\alpha_{SiSi}$, $\alpha_{AlAl}$, $\alpha_{SiAl}$, and $\alpha_{AlSi}$ for the second shell is shown in Fig.~\ref{temp2ndshell} and used to investigate at which temperature the phase separation is manifested in the samples across all compositions. At elevated temperatures, the behavior of $\alpha_{SiSi}$ and $\alpha_{AlAl}$ of the second neighboring shell is similar to that at 300~K, with only the intensity being different. On the other hand, and at high temperatures, it is evident from the graph that the WC parameter of the $\alpha_{SiAl}$, and $\alpha_{AlSi}$ tends towards lower values, negative in case of $\alpha_{SiAl}$, and positive ones in case of $\alpha_{AlSi}$, which indicate that in the liquid phase, SiO$_4$ attract AlO$_n$. The WC parameter becomes increasingly repulsive as the temperature decreases, particularly around the glass transition temperature. This suggests that aluminum polyhedra prefer to be surrounded by other aluminum polyhedra. This tendency promotes immiscibility in the supercooled liquid phase, leading to phase separation. The WC parameter shows that the sample exhibits a strong tendency towards demixing in the low-temperature region, as indicated by the higher positive values.

The findings presented in this study shed light on the origins of the nanoscale phase separation in binary aluminosilicate glasses. The repulsion between SiO$_4$ and AlO$_n$ polyhedra is shown to be the origin of the phase separation in aluminosilicate glasses at the atomic scale. In agreement with previous experimental studies~\cite{macdowell1969, risbud1977, KLUG1987, Wilke2022, Wilke2023}, our analysis showed that the composition at which the phase separation is maximized in binary AS glasses is around 50 mol\% of Al$_2$O$_3$.
The underlying hypothesis of this discussion is that phase separation refers to the redistribution of atoms from a homogeneous mixture to form distinct regions with different local atomic arrangements. This process occurs to minimize the energy of the system and to accommodate the differences in the local potential energy between different regions of the system. We computed the atomic potential energy for Si and Al to verify this hypothesis. The atomic potential energies ($<U_i>$) were calculated for both Al and Si atoms. Subsequently, the variance of the potential energy per atom for each type of atom was calculated using the formula: $\sigma^2_i$ = (1/N$_i$)(U$_i$-$<U_i>$)$^2$, where N$_i$ represents the number of atoms of type $i$ and U$_i$ is the potential energy of atoms of type $i$. 
\begin{figure}[htb!]
\centering
\includegraphics[width=\columnwidth]{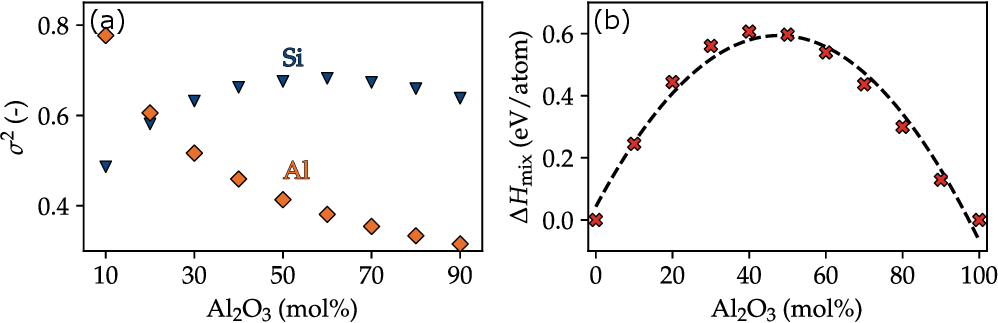}
\caption{(a) The variance of atomic potential energy for Si and Al atoms. (b) Variation of the mixing enthalpy as a function of alumina content at 300~K. The lines in (b) are to guide the eye (For interpretation of the colors in the figure(s), the reader is referred to the web version of this article).}
\label{Fig:energy}
\end{figure}
In Fig.~\ref{Fig:energy}(a), the variance obtained from the atomic potential energy distribution for both Si and Al atoms is shown. This shows that as the concentration of Al$_2$O$_3$ increases, there is a corresponding non-linear decrease in the variance for Al with a change of slop around 50 mol\% of Al$_2$O$_3$, indicating the local environment of Al becomes more homogeneous with increasing Al$_2$O$_3$ content.

On the other hand, this observation holds physical significance: the environment of SiO$_4$ tetrahedra is more homogeneous at low alumina content. With increasing Al$_2$O$_3$, the values of the variance increase to reach a maximum of around 50 mol\% of Al$_2$O$_3$ and then decrease again. This also indicates that the highest deviation from the mean potential energy of Si atoms occurs when the Al$_2$O$_3$ content is approximately 50 mol\%. This might be associated with the emergence of zones characterized by elevated potential energy. Such regions could signify areas where atomic interactions are less favorable, forming distinct phases. Thus, the phase separation in aluminosilicate glasses is due to a non-random distribution of the SiO$_4$ and AlO$_n$ polyhedra, caused by the increasing attraction between SiO$_4$ tetrahedra and repulsion between SiO$_4$ tetrahedra and AlO$_n$ polyhedra. These interactions between the network and the former polyhedra lead to the formation of distinct phases to minimize the system energy. This behavior is well captured by the positive mixing enthalpy $\Delta H_{\rm mix}$ (detail of the calculation of $\Delta H_{\rm mix}$ is given in the SM) obtained from the MD simulations, as shown in Fig.~\ref{Fig:energy}(b). For all compositions, the mixing enthalpy is positive, indicating the presence of a demixing in the glass with its maximum being around 50 mol\% of Al$_2$O$_3$. 

In summary, the current study has contributed to a clearer understanding of the origins of nanoscale phase separation in aluminosilicate glasses. We showed that compositions containing approximately 50 mol\% of alumina exhibited the most pronounced phase separation. The origins of the nanoscale phase separation are caused by a preference for SiO$_4$ and AlO$_n$ polyhedra to cluster starting from the second neighboring shells, ultimately leading to the formation of distinct, homogeneous phases locally. 
This was demonstrated through the chemical order analysis at larger coordination shells using the WC parameter. Moreover, we showed that the phase separation is particularly pronounced in the vicinity of the glass transition temperature and continues to be notable at lower temperatures.

\section*{Conflicts of interest}
There are no conflicts to declare.

\section*{Author Contributions}
H.K. and A.A. performed the simulations and analysis and wrote the original draft. A.H. contributed to the interpretation of the results. A.A. designed and supervised the research and analysis. All authors contributed to the editing of the final manuscript.

\end{document}


\title{Supplementary Materials to:\\The origin of phase separation in binary aluminosilicate glasses}

\author{Houssam Kharouji}
\affiliation{LEM3,Labex Damas, Université de Lorraine, Metz, 57070, France}
\affiliation{LS2ME, Facult\'{e} Polydisciplinaire Khouribga, Sultan Moulay Slimane University of Beni Mellal, B.P 145, 25000 Khouribga, Morocco}
\author{Abdellatif Hasnaoui}
\affiliation{LS2ME, Facult\'{e} Polydisciplinaire Khouribga, Sultan Moulay Slimane University of Beni Mellal, B.P 145, 25000 Khouribga, Morocco}
\author{Achraf Atila}
\email{achraf.atila@uni-saarland.de; achraf.atila@gmail.com}
\affiliation{Department of Material Science and Engineering, Saarland University, Saarbr\"{u}cken, 66123, Germany}

\date{\today}
\maketitle 

\section{Methods}
\subsection{Potential model}
The interactions between atoms were modeled using the Born-Mayer-Huggins potential function. This model includes long-range Coulomb interactions (first term), short-range exponential function (second term), and dispersion energy that takes into account the dipole-dipole interactions (third term), as given by eq. \ref{potential}.
\begin{equation}
\label{potential}
U_{ij}(r) = \frac{q_iq_j}{4\pi \epsilon_0r} + A_{ij} exp\left(\frac{\sigma_{ij} - r}{\rho_{ij}}\right) - \frac{C_{ij}}{r^6},
\end{equation}
where $i$ and $j$ are atoms (Si, O, or Al), r$_{ij}$ is the distance between atoms $i$ and $j$, q$_i$ is the effective charge of the atom $i$, and $A_{ij}$, $\sigma_{ij}$, $\rho_{ij}$, and $C_{ij}$ are potential parameters. For the parameters, we use the ones provided by Bouhadja~\textit{et al.}~\cite{Bouhadja2013}, which were shown well to describe the atomic structure and properties of aluminosilicate glasses and melts \cite{atila2019alumina, Tang2023}. 
\subsection{Glass preparation}
Nine binary aluminosilicate glasses (SiO$_2$)$_{(1-x)}$--(Al$_2$O$_3$)$_{x}$, $x = 0.1 - 0.9$), were simulated using molecular dynamics simulations using LAMMPS package~\cite{Thompson2022}. For all samples, we randomly distributed around a million atoms in periodic cubic simulation boxes with a volume chosen to give a density larger than the experimental glass density by 5~\%. No unrealistic overlap between atoms was allowed. The samples were equilibrated at high temperature ($T$ = 5000 K) for 20 ps in the NVT (constant number of atoms, volume, and temperature) ensemble and for 2 ns in the NPT ensemble (constant number of atoms, pressure, and temperature) with the external pressure set to 0 MPa. The obtained melts were subsequently cooled to room temperature ($T$ = 300 K) using a cooling rate of 1 K/ps while keeping the pressure around zero MPa. Then at 300 K the glasses were relaxed in the NPT ensemble for 2 ns and another 2 ns in the NVT ensemble. The equations of motion were evaluated using the velocity-Verlet algorithm with a timestep of 2 fs. The temperature and pressure were maintained using the Nos\'e--Hoover chain thermostat and barostat with 10 chains each and coupling factors of 200 fs and 2000 fs for the thermostat and barostat, respectively.

\subsection{Calculation of local density}
\begin{figure}[h!]
\centering
\includegraphics[width=\columnwidth]{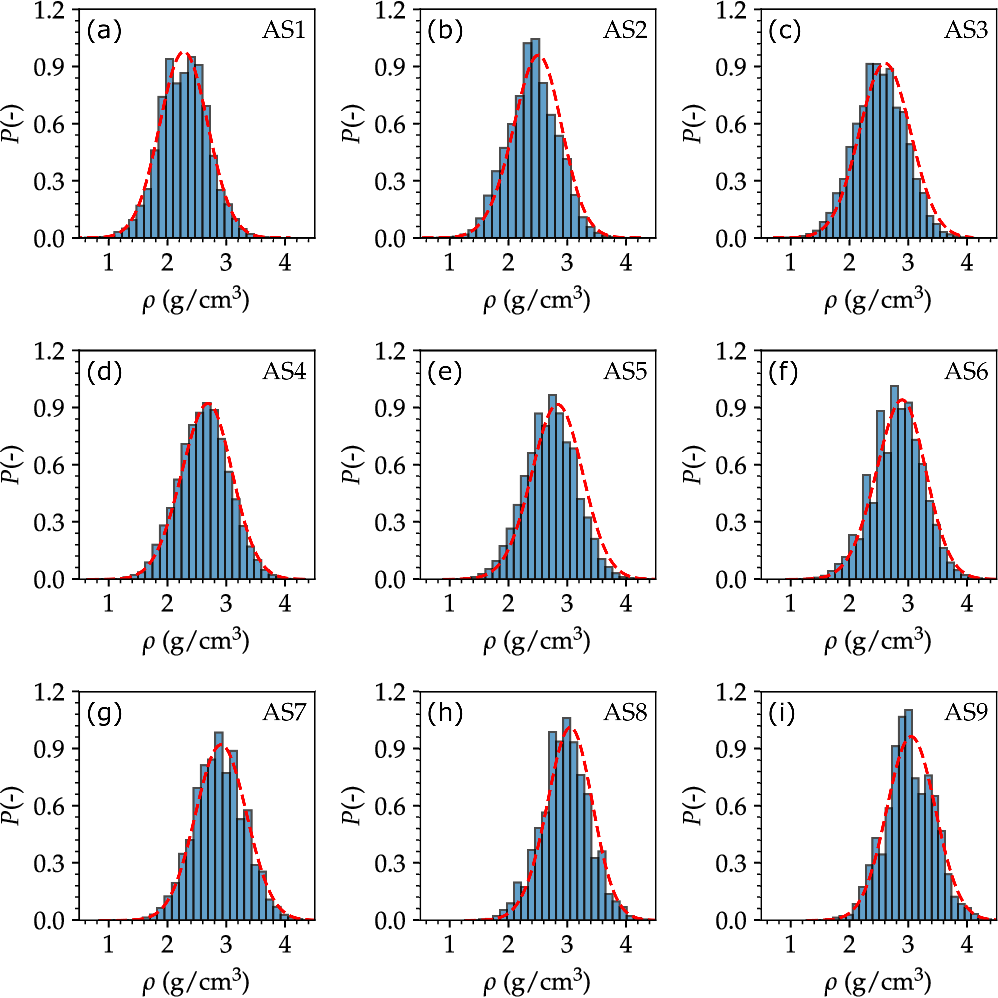}
\caption{The distribution of the local densities for all samples at 300~K. The red dashed lines correspond to skewed Gaussian fits.}
\label{Density}
\end{figure}

The local mass density was calculated by partitioning the simulation box into cubic voxels of size $10\times10\times10$~\text{\AA$^3$}, and the local mass density was computed within each voxel. 
The extracted densities are plotted as a histogram (See Fig.~\ref{Density}) and fitted using a skewed Gaussian distribution~\cite{Sukhomlinov2017} (See eq.~\ref{Gaussian}).
\begin{equation}
f(\rho_i ,\mu ,\sigma ,\xi) = \frac{e^{[-(\rho_i-\mu)^2/2\sigma^2]}}{\sigma\sqrt{2\pi}}\left\{1 + \erf \left(\xi\frac{(\rho_i - \mu)}{\sigma\sqrt{2}}\right) \right\},
\label{Gaussian}
\end{equation}
where $\rho_i$ is the local density, $\mu$, $\sigma$, and $\xi$ are the coefficients determining the first two moments of the local density distribution. The skewness can then be extracted from the fitting parameters as in eq.~\ref{skewness}
\begin{equation}
S = \frac{4-\pi}{2} \frac{\tilde{\xi}^3}{(\pi/2 - \tilde{\xi}^2)^{3/2}},
\label{skewness}
\end{equation}
with $\tilde{\xi} = \xi/\sqrt{1 + \xi^2}$.

\subsection{Enthalpy of mixing}
The atomic enthalpy of mixing $\Delta H_{\rm mix}$ reveals whether a glass of a given composition will tend to phase separate or form a homogeneous phase. It can be obtained from MD simulations using the expression:
\begin{equation}
\Delta H_{\rm mix} = H_{\rm AS} - (1-x)H_{\rm SiO_2} - xH_{\rm Al_2O_3},
\end{equation}
where $H_{\rm AS}$ is the enthalpy calculated from the MD simulations for the glass AS, $H_{\rm SiO_2}$, and $H_{\rm Al_2O_3}$ are the atomic enthalpies of the end-member compositions obtained using the same melt-quench process as the AS glasses.

\section{Structure of the glasses}

\subsection{Radial distribution function and coordination numbers}
The radial distribution function (RDF)  provides insights into the local atomic arrangement within the system as well as the structure at extended scales. It characterizes the probability of finding a particle $i$ from a given particle $j$ at a distance $r$. Otherwise, the coordination number (CN) can be estimated by integrating the area under the first peak of the RDF. The CN can be expressed as follows:

\begin{equation}
\label{coordination}
    N_{ij}(r_{c}) = 4\pi \rho \int_0 ^{r_{c}} g_{ij}(r)r^2 dr,
\end{equation}
where $r_c$ is the cutoff value and $\rho$ is the atomic density.
\begin{figure}[h!]
\centering
\includegraphics[width=\columnwidth]{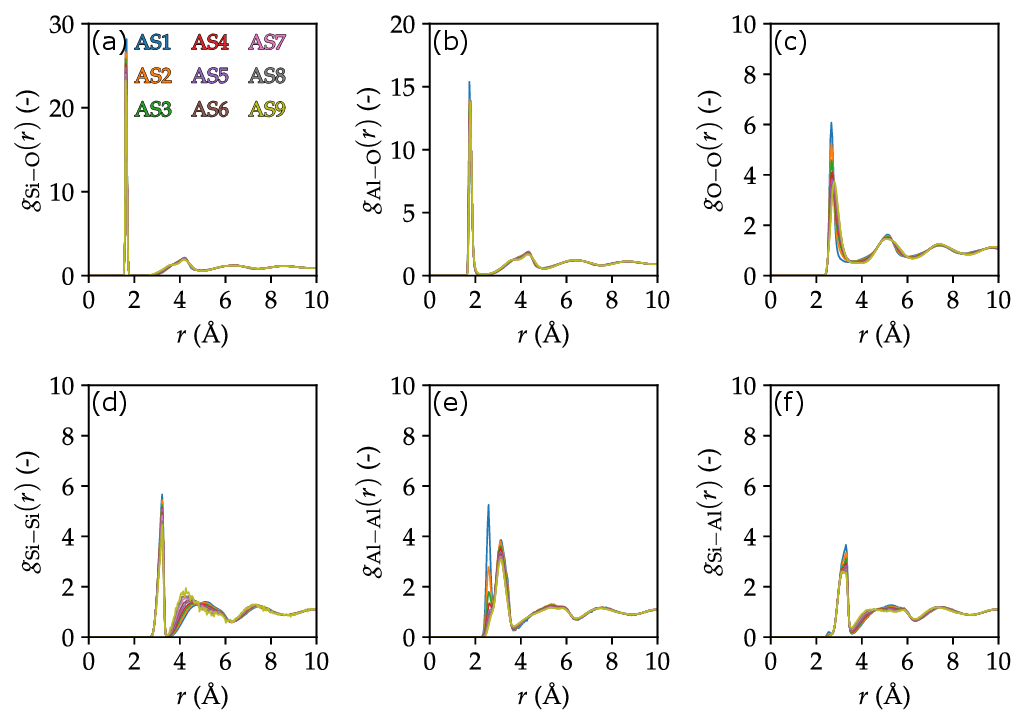}
\caption{Radial distribution functions for all glasses at 300~K. (a) for Si–O, (b) Al-O, (c) O-O, (d) Si-Si, (e) Al-Al, and (f) Si-Al.}
\label{RDF}
\end{figure}
Fig.~\ref{RDF}(a) and Fig.~\ref{RDF}(b) illustrate the first peaks of Si-O and Al-O RDFs, respectively. The maxima position of each RDF curve provides valuable information concerning the average distance between the Si/Al atom and its first oxygen neighbors, which corresponds to the bond length. We notice that the first peaks of the Si-O and Al-O RDFs exhibit maxima at distances approximately 1.62~\text{\AA} and 1.78~\text{\AA}, respectively, which are in good agreement with the experiment data reported in the literature~\cite{weber2008, Hanada1982}. As the alumina content increases, the (RDFs) for Si-O and Al-O pairs maintain their peak maxima positions. Nevertheless, a pronounced decrease was shown in the peak intensity. Hence, the average coordination numbers within the first coordination shell are determined by integrating the partial radial distribution functions up to a cutoff radius of 2.1 and 2.3~\text{\AA} for Si-O and Al-O atom pairs, respectively. The findings indicate that the coordination number of silicon remains consistently at 4 across all compositions, while the coordination number of aluminum rises slightly as the alumina content increases (see table). The evolution of the first peaks of the O-O the partial distribution function is presented in Fig.~\ref{RDF}c.
The results show that as the Al$_2$O$_3$ content increases, the position of the first peak shifts towards higher distance values. The O-O bond length varies from 2.66~\text{\AA} for AS1 to 2.78~\text{\AA} for AS9. Subsequently, a noticeable reduction in the intensity of the first peak occurs, accompanied by an increase in its width indicating the presence of different polyhedra in the glasses AlO$_n$ ($n = $4, 5, 6).

\begin{table}
\caption{\label{tab2}Short-range structural parameters of glasses obtained from molecular dynamics at 300~K. The cutoffs to calculate the mean coordination numbers were
set to 2.0~\text{\AA} for the Si–O, 2.3~\text{\AA} for Al-O, and 3.5~\text{\AA} for O-O, respectively.}
\begin{ruledtabular}
\begin{tabular}{ccccccc}
\hline
\multicolumn{1}{c}{Glass ID} & \multicolumn{2}{c}{Si-O} & \multicolumn{2}{c}{Al-O} & \multicolumn{2}{c}{O-O} \\
\cline{2-3} \cline{4-5} \cline{6-7} 
 & $r_{ij}$ & $N_{ij}$ & $r_{ij}$ & $N_{ij}$ & $r_{ij}$ & $N_{ij}$ \\ 
\colrule
AS1 & 1.66 & 4.00 & 1.74 & 3.98 & 2.66 &  7.66 \\
AS2 & 1.62 & 4.00 & 1.74 & 4.04 & 2.66 &  8.68 \\
AS3 & 1.62 & 4.00 & 1.78 & 4.08 & 2.66 &  9.91 \\
AS4 & 1.62 & 4.00 & 1.78 & 4.13 & 2.70 &  10.53 \\
AS5 & 1.62 & 4.00 & 1.78 & 4.15 & 2.70 &  10.90 \\
AS6 & 1.62 & 4.00 & 1.78 & 4.19 & 2.70 &  11.62 \\
AS7 & 1.62 & 4.00 & 1.78 & 4.23 & 2.74 &  12.51 \\
AS8 & 1.62 & 4.00 & 1.78 & 4.26 & 2.74 &  12.54 \\
AS9 & 1.62 & 4.00 & 1.78 & 4.29 & 2.78 &  13.17 \\
\end{tabular}
\end{ruledtabular}
\end{table}

\subsection{Bond angle analysis}
For a more comprehensive exploration of the spatial short-range structure around silicon and aluminum, we rely on the calculation of angular distribution function (ADF). This approach allows us to assess the spread of bond angles formed between a central atom $j$ and its adjoining species $i$ and $k$, offering valuable insights into the local atomic arrangement. The bond angle is estimated as a function of the distance between atoms as follows~\cite{atila2019atomistic}:
\begin{equation}
\theta_{ijk} = \arccos\left(\frac{r_{ij}^2 + r_{ik}^2 - r_{jk}^2}{2r_{ij}^2 r_{ik}^2}\right).
\end{equation}
\begin{figure}[h!]
\centering
\includegraphics[width=\columnwidth]{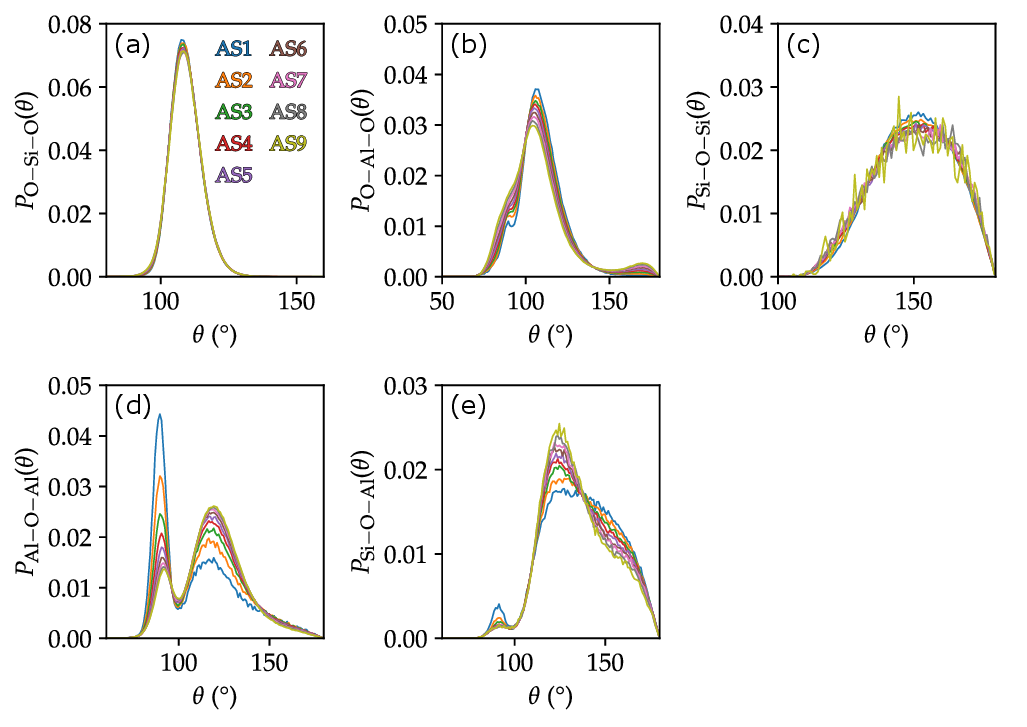}
\caption{Bond angles distributions obtained from MD simulations at 300 K. (a) O-Si-O, (b) O-Al-O, (c) Si-O-Si, (d) Al-O-Al, and (e) Si-O-Al.}
\label{ADF}
\end{figure}

Figure~\ref{ADF} displays the distributions of bond angles. The O-Si-O BAD presents the distribution of the angles inside SiO$_4$ tetrahedra. The average O-Si–O angle centers around 108.4\degree, which closely approximates the ideal value found in a perfectly tetrahedral geometry, which is about 109.5\degree. Notably, the addition of Al$_2$O$_3$ content does not significantly affect the O–Si–O bond angle. On the other hand, in Fig.~\ref{ADF}(b), it is evident that the distribution of O-Al-O bond angles is wider compared to that of O-Si-O, with its central tendency at 105.8\degree shifting towards lower angle values and while the peak intensity decreases as the alumina content increases. Fig.~\ref{ADF}(c) illustrates the distributions of Si-O-Si bond angles, depicting the linkage angles between SiO$_4$-SiO$_4$ tetrahedra. Across all systems, the Si-O-Si bond angle consistently centers around 147\degree. In the Al-O-Al angular distribution function (Fig.~\ref{ADF}(d)), two peaks are observed. The first peak, approximately at 90\degree, signifies the existence of AlO$_5$ and AlO$_6$ polyhedra. Conversely, additional peaks around 120\degree are observed, attributed to the presence of AlO$_4$ tetrahedra. Fig.~\ref{ADF}(e) depicts the Si-O-Al bond angle distribution, which has a main peak of around 140\degree with its intensity increasing with composition and a small peal around 90\degree.\\

\begin{figure}[h!]
\centering
\includegraphics[width=\columnwidth]{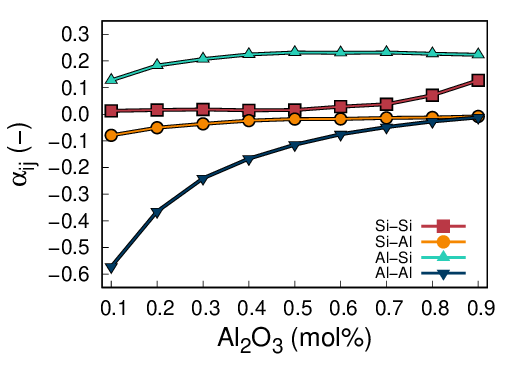}
\caption{WC parameter $\alpha_{\rm ij}$ up to the first shell at 300~K as a function of the alumina content.}
\label{WCparam}
\end{figure}
\begin{figure}[h!]
\centering
\includegraphics[width=\columnwidth]{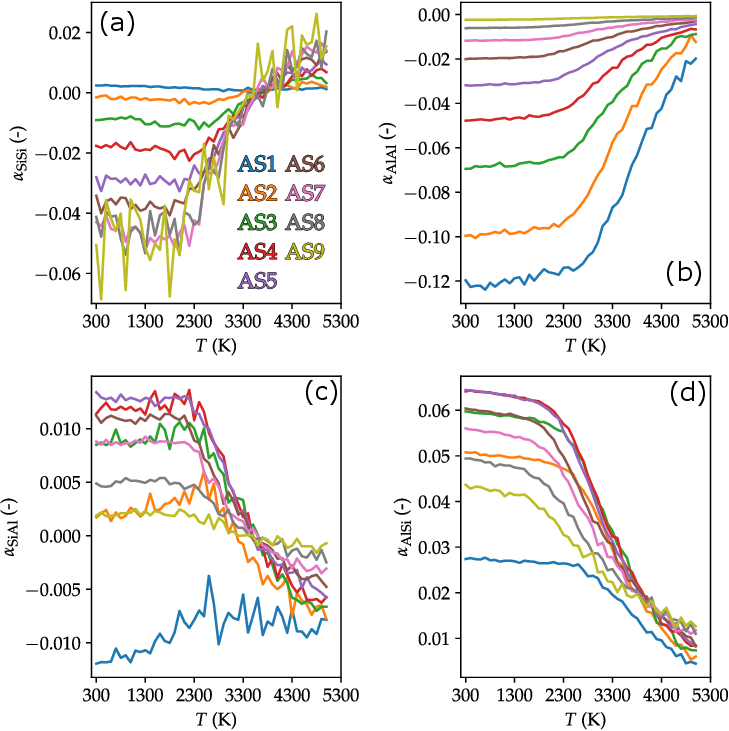}
\caption{WC parameter $\alpha_{\rm ij}$ of the second coordination shells as a function of the temperature for all samples (a) $\alpha_{\rm SiSi}$, (b) $\alpha_{\rm AlAl}$, (c) $\alpha_{\rm SiAl}$, and (d) $\alpha_{\rm AlSi}$.}
\label{WC_temp_secondshell}
\end{figure}

\newpage

\bibliography{main}